\documentclass[%
 reprint,
 amsmath,amssymb,
 aps,
]{revtex4-2}

\usepackage{graphicx}
\usepackage{dcolumn}
\usepackage{bm}

\begin{document}

\preprint{APS/123-QED}

\title{On thermodynamic stability and ensemble equivalence at coexistence}

\author{Jozismar Rodrigues Alves}
 \email{jozismar.alves@alumni.usp.br}
\author{Vera Bohomoletz Henriques}%
 \email{vera@if.usp.br}
\affiliation{%
 Instituto de Física - Universidade de São Paulo.
}%




\date{\today}

\begin{abstract}
Numerical simulations of systems at coexistence are known to yield unstable fields in some regions of the density parameters, as well as inequivalence of ensembles. The ‘Van der Waals-like’ loops are attributed to effects of the interface between the coexisting phases, but the question of convexity remains unresolved. We recover the theory developed by Hill in the 1960's, and adapt it to include the interface free energy. Our adapted theory is verified through carefully planned simulations, and give a thermodynamic description of the interface behaviour
inside the coexistence region which restores the proper convexity of the thermodynamic potentials, as well as yields convergence of grandcanonical and canonical results. As a bonus, our interpretation allows direct calculation of surface tension in very good accordance with Onsager's analytic prediction. 
\end{abstract}

\maketitle


The XIXth century experimental study of the coexisting gas-liquid phases \cite{Andrews} showed that the system pressure-volume isotherms present a horizontal plateau which signals the presence of phase separation. The first successful theory of Andrews \cite{Andrews} results was proposed by Van der Waals \cite{vanderWaals}: it had
the advantage of presenting two densities for the same pressure, as in coexistence. However, instead of the plateau, it presented loops. Loops are present in all "classical" mean-field theories \cite{Callen, Huang} for phase transitions, 
instability arising from a
mathematical treatment which does not include description of phase separation, since
the system is treated as homogeneous all the way through the coexistence region.

\begin{figure}
\includegraphics[width=9cm, height=6cm]{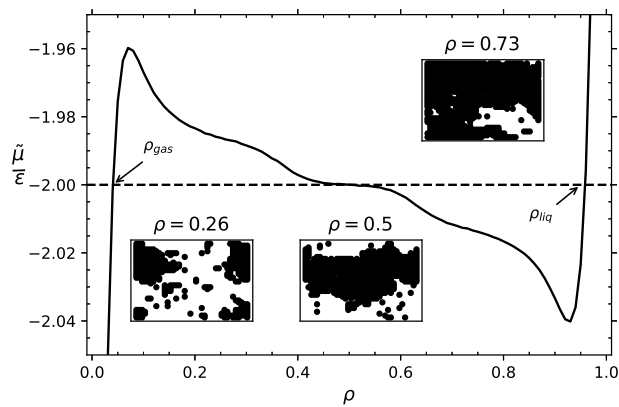}
 \caption{\label{figure1}Lattice gas chemical potential vs. density isotherm below criticality obtained from MC simulations in the $(T, V, N)$ canonical ensemble. Insets correspond to system snapshots for densities in the +-loop, plateau and in the -loop, respectively. The dashed line is an extension of the plateau region and its intersections with the isotherm correspond to the phase coexistence densities (see text). Data are for $L=40$ and temparature $t=0.5$.}
\end{figure}

In the absence of exact treatment for most systems, numerical simulations constitute the only resource for exploring thermodynamic properties of many statistical or molecular models which take into account correlations. However, it has been known for a few decades \cite{furukawa1982two,BenjaminAndBinder, hill1987statistical}, that the numerical simulations of molecular or statistical models in fixed density ensembles present ''Van der Waals-like'' loops on the sides of the plateaus (see Figure \ref{figure1}). The second law of Thermodynamics imposes strict conditions on the thermodynamic state variables. Fields must present a definite
convexity in relation to their conjugate densities, thus loops would imply violation of thermodynamic stability. Several studies have attributed loops to interface effects in finite systems, which should disappear in the thermodynamic limit \cite{hill1987statistical,hill1987statistical}. The
de-escalating of loops with size has been indicated very clearly in several numerical studies \cite{BenjaminAndBinder}. 
As to fixed field ensembles, instead of loops one sees a region of "hystheresis", wich has been interpreted in terms of metastability.
Thus, numerical simulations of phase-separating systems in the coexistence region seem to yield both thermodynamic instability as well as the nonequivalence of ensembles. Both features have been reputed to be a consequence of the finiteness of the simulated systems. 

In this paper, we recover a theory proposed by Hill \cite{hill1987statistical}  for the thermostatistics of different ensembles of systems at coexistence, presented by the author at the dawn of numerical studies of statistical models. We add a new interpretation to Hill's approach, by introducing the thermodynamics of the interface between coexisting phases. Our proposal points to the recovery of proper thermodynamic stability, as well as to the equivalence of ensembles, as we illustrate for the case of the square lattice fluid model.

What is the precise origin of loops in numerical simulations? In the 1960’s, Hill \cite{hill1987statistical} presented a theory aimed at answering this question, which we describe briefly. Let $Y$  represent a thermodynamic field and $X$ the respective conjugate density. At a particular density $X$, a system at coexistence presents
two different densities, say $X_1$ and $X_2$, which represent the densities of the two coexisting phases. If one runs simulations for fixed conjugate field $ Y$, density $X$  fluctuate
around the two values $X_1$ and $X_2$. Thus, if one plots the distribution probability density $P(X|Y$), one obtains a two-peak function, with peaks centered on $X_1$ and $X_2$, separated by a minimum at $X_{min}$.

For clearness of argument, let us consider a simple fluid at the gas-liquid coexistence.
The grand-canonical probability of finding $N$ particles in the system at $(T, V, \mu)$ is given by
\begin{equation}
 P(N; T, V, \mu)= \frac{e^{\beta \mu N}Z(T, V, N)}{\Xi(T, V, \mu)}
 \label{equation1}
\end{equation}
where $Z(T, V, N)$ and $\Xi(T, V, \mu)$ are, respectively, the canonical and the grand canonical partition functions of the system.

According to Hill's proposal, the maximum of the distribution $\left[\frac{dP(N; T, V, \mu)}{dN}\right]_{N_{gas}, N_{min}, N_{liq}}=0$ should yield the chemical potential in the canonical ensemble:

\begin{equation}
 \beta \mu +\frac{1}{Z(T, V, N)}\frac{\partial Z(T, V, N)}{\partial N}=0
 \label{Equation2}
\end{equation}
 in accordance with the corresponding equation of state in the Helmholtz free-energy representation 
 
 \begin{equation}
 \left(\frac{\partial F}{\partial N} \right)_{T, V} = \mu (T, V, N)
 \label{Equation3}
\end{equation}.

Liquid-gas coexistence would be signaled by a two-peaked distribution of particle number $N$ for the system in the chemical potential bath, thus yielding
three values for particle density, corresponding to $N_{gas}$, $N_{min}$, $N_{liq}$, at the same
chemical potential $\mu$. At coexistence, probability distribution $p(N)$ should present two maxima of the same height, while the minimum would run over a continuous set of values of the abcissa $N$.

We argue that the theory proposed by Hill \cite{hill1987statistical} is in fact correct for the homogeneous phases, but must be
modified in its application to the coexistence region, because of the additional
dependence of free-energy on the thermodynamics of the interface between the two
phases. For a simple fluid, the free-energy for the system at coexistence, $F_{cx}$, must be written as

\begin{equation}
 F_{cx}(T, V, N) = F_{bulk}(T, V, N) + F_{int}(T,V,N)
 \label{Equation4}
\end{equation}
with the free-energy of the interface given by $F_{int} = \gamma(T) A_{int}(T,V,N)$, 
where $\gamma$ describes surface tension, while $A_{int}$  is the area of the interface \cite{RowlinsonWidom, Widom}.
As for the equation
of state related to variation of the number of particles, we have:

\begin{equation}
 \left( \frac{\partial F_{cx}}{\partial N} \right)_{T, V} = \mu_{coex}(T, V, N) + \gamma \left( \frac{\partial A_{int}}{\partial N} \right)_{T, V},
 \label{Equation5}
\end{equation}
where chemical potential is obtained from the bulk free-energy as

\begin{equation}
 \mu(T, V, N) = \left( \frac{\partial F_{bulk}}{\partial N} \right)_{T, V} = \left( \frac{\partial F_{cx}}{\partial N} \right)_{T, V, A_{int}(T,V,N)} . 
 \label{Equation6}
\end{equation}

Thus it is clear that computing the partial derivative of the free energy with respect to particle number, at fixed temperature and volume, does not yield the chemical potential (Eq.\ref{Equation3}, but, instead, the sum of the chemical potential with a term proportional to the variation of the interface area with particle number (Eq.\ref{Equation6}).

This special form of the free-energy
at coexistence must be associated to a modified grand-partition function which takes into account the interface area $A_{int}$. Thus, the probability distribution for the number of particles in the grand-canonical ensemble (Eq. \ref{equation1}) in the coexistence region  must be rewritten as 

\begin{equation}
 P(N; T, V, \mu) = \frac{e^{\beta(\mu N + \gamma(T) A_{int})} Z(T, V, N)} {\Xi(T, V, \mu)}
 \label{equation7}
\end{equation}
Extrema with respect to particle number $N$ of the the
distribution function at coexistence $P(N; T, V, \mu)$ are given by:

\begin{equation}
 \beta \mu + \gamma \left( \frac{\partial A_{int}}{\partial N} \right)_{T, V}+\frac{1}{Z(T, V, N)}\frac{\partial Z(T, V, N)}{\partial N}=0
 \label{Equation8}
\end{equation}.
The corresponding equation of state is then given by Eq. \ref{Equation6}, instead of Eq. \ref{Equation3}.

What are the implications of our proposal on the interpretation of simulation results? 

In canonical ensemble 
simulations, the numerical chemical potential is obtained through Widom's insertion method  \cite{Widom}. From discretization of Eq.\ref{Equation3} for homogeneous systems, a numerical chemical potential $\mu_{num}$ may be shown to be a function of an average of
the Boltzman factor of the energy added to the system upon insertion of a 'virtual' particle: 
\begin{widetext}
\begin{equation}
\mu_{num}(T, V, N) \equiv - k_BT \ln  \frac{Z(T,V,N+1)}{Z(T,V,N)}  =  k_BT \ln \left[ \frac{\rho}{\langle \exp \left(\frac{e_{additionalparticle}}{k_BT}\right)\rangle_{T, V, N}}\right].
 \label{Equation9}
\end{equation}
\end{widetext}

As we discussed above, Eq. \ref{Equation3}, and thus also Eq. \ref{Equation9},  in the case of simulations, yield the chemical potential only for the case of homogeneous phases. In the coexistence region, numerical simulations in the canonical ensemble yield non-homogeneous phases, and the true chemical potential $\mu$ is no longer given by the discretized partial derivative of the canonical free-energy. Rather,  the latter  is really a combination of chemical potential and a term proportional to the gradient of the area of the interface. Thus, at coexistence, Widom's insertion (\ref{Equation9}) yields a pseudo-chemical potential, which we design $\tilde{\mu}$, related to the true chemical potential $\mu$ through 

\begin{equation}
 \tilde{\mu}(T, V, N) = \mu(T, V, N) + \gamma \left( \frac{\partial {\tilde{A}_{int}}}{\partial N} \right)_{T, V}.
 \label{Equation10}
 \end{equation}
 
  Note that at coexistence, $\tilde{\mu}$ must rise above the true chemical potential $\mu$, as the area of the interface increases with particle number $N$, whereas the opposite is true if adding particles brings down the interface area. Also, if interface area is maintained constant, as particles are added to the system, the pseudo chemical potential equals the true chemical potential,$\tilde{\mu}$ = $\mu$.

In order to check our modified Hill's theory, we have carried out very careful simulations
for the lattice fluid on the square lattice, under periodic boundary conditions. For the chosen model, analytical results by Onsager are
available \cite{Onsager1944}, which serve as good test on the theory. Our procedure is as follows: (i) in the grand-canonical ensemble $(T, V, \mu)$, we calculate the full particle density probability distributions $P(N)$, whose extrema yield precise pseudo chemical-potential {\textit{vs}} density $rho$ relations; (ii) we obtain the pseudo chemical potential isotherms directly from usual canonical ensemble $(T,V, N)$ simulations, through eq \ref{Equation9}; (iii) Results from the two different ensembles are compared.

\begin{figure}
\includegraphics[width=9cm, height=6cm]{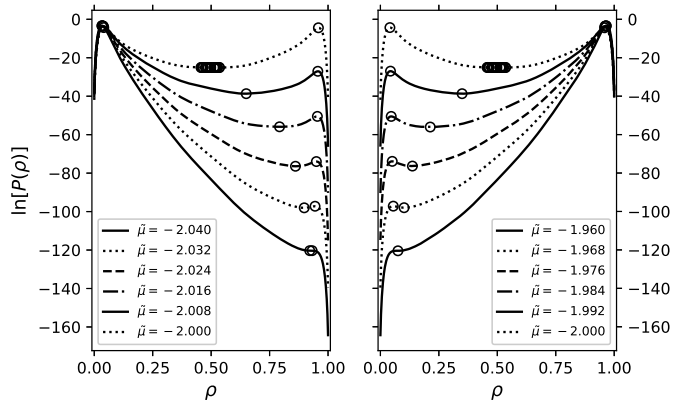}
 \caption{\label{figure2}Probability distribution function $P(N; T, V, \mu)$ for different values of the chemical potential $\mu$, obtained from $(T, V, \mu)$ grand-canonical ensemble. The left curve displays $P(N)$ for decreasing values of the chemical potential from coexistence and below, while the right curve displays $P(N)$ for increasing chemical potential values above coexistence. Circles indicate extrema of the distributions. Data are for $L=40$ and temperature $t=0.5$.}
\end{figure}

In the {\textit {grand-canonical}} ensemble $(T,V, \mu)$ description, the probability distribution function $P(N; T, V, \mu$), at different chemical potentials, is obtained via a combination of multicanonical
and Wang-Landau techniques, as proposed in references \cite{Berg1991, Berg1992, WangLandau2001}.  The use of these techniques is central to our results, because they allow us to probe events of very low probability, in between the gas and liquid densities, thus yielding the full $\mu$ vs $\rho$ isotherms. The N probability distributions are represented in Fig \ref{figure2}, for
different values of the pseudo chemical potential $\tilde {\mu}$ (see equation \ref{equation7}). It can be seen that there are two equal peaks for the
coexistence chemical potential $\mu/\epsilon = -2$ (corresponding to zero magnetic field for the Ising model). As
the chemical potentials rises above the coexistence value, the small density peak
diminishes while the large density peak increases. The opposite is true if the chemical
potential is lowered below its coexistence value. 

In the {\textit{canonical}} $(T,V,N)$ ensemble, the usual Metropolis algorithm \cite{BenjaminAndBinder} was used to obtain $\tilde{\mu}$ from the equation of state represented by Eq. \ref{Equation10}. 

\begin{figure}
\includegraphics[width=8.5cm, height=6cm]{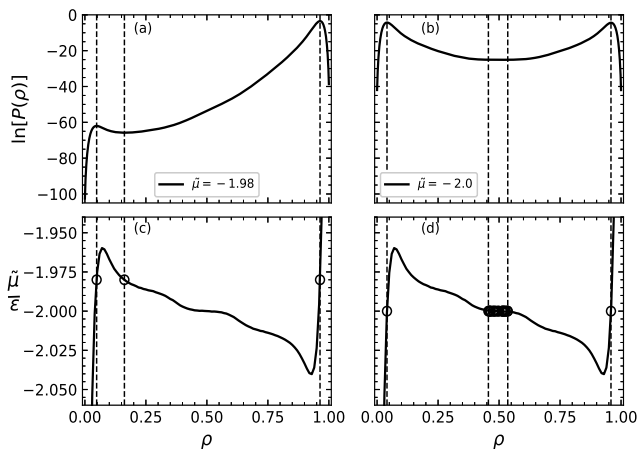}
\caption{\label{figure3}Comparison between results obtained from $(T, V, N)$ and $(T, V, \mu)$ simulations. Figures (a) and (b) display distribution probability functions $P(N; T, V, \mu)$ for $\mu=-2.0$ and for $\mu=-1.98$, respectively. In Figures (c) and (d), the same chemical potential isotherm (for $t=0.5$) obtained from $(T, V, N)$ simulations is displayed, in order to facilitate comparison with the $(T, V, \mu)$ results. The dashed vertical lines connect the extreme points of $P(N$) at fixed $\mu$ with the corresponding densities on the chemical potential $(T, V, N$) isotherm. Note the correspondence between chemical potentials in each of the two sets of data. Data points are for $L=40$ and temperature $t=0.5$.} 
\end{figure}

Our numerical results for both ensembles are illustrated in the bottom Fig. \ref{figure3}. Upper plots show grand-canonical probability distributions $P(N)$ and lower plots display canonical pseudo-chemical potential $\tilde \mu$. The results in the different ensembles may be compared in the folowing terms: (i) on the left pannel, extrema of the grand-canonical distribution function $P(N)$ for a given particular $\tilde \mu$ are in correspondence with the density values at the same $\tilde \mu$ of the canonical isotherms in the loop region; (ii) a small plateau for the canonical $\tilde{\mu}$, seen at intermediate densities, is in correspondence with the set of minima of the $P(N)$ probability distribution,
while a 'plus' loop is seen near low densities and a 'minus' loop is seen near large densities. 
Let us analyse these results in the light of state equation \ref{Equation5} (or Eq. \ref{Equation10}). In the plateau region of intermediate densities, pseudo and true chemical potential converge, and interface area should be stationary under increase in particle number. This means we can identify the coexistence
potential with the plateau value, and extract the coexisting densities from the
intersection points of the extension of the plateau with the $\tilde{\mu}$ curve (see Fig.\ref{figure1}). Note that the plateau occurs at the value predicted by the exact Onsager theory, $\frac{\mu_{cx}}{\epsilon}=-2$. 
How can we interpret the effects of variation in interface area?  As one goes over the lower density limit for coexistence, an interface area appears and increases as particles are added, and $\tilde{\mu} > \mu$. In a planar system, a circular dense phase is formed, but as particle number increases, it deforms towards a rectangular strip accross the lattice, because of periodic boundary conditions. The interface area of the rectangular strip is independent of its width, and thus remains constant as new particles are inserted into the system. This is true up to the point, at higher densities, at which the strip initiates deformation, as the low density phase contracts to a circle whose perimeter decreases as particle number decreases. In this region, interface area decreases, and pseudo potential falls below true chemical potential.

\begin{figure}
\includegraphics[width=8cm, height=5cm]{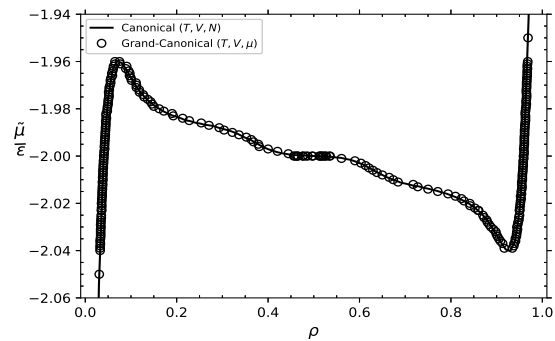}
\caption{\label{figure4}Comparison between thermodynamics obtained from different ensembles. Results for the pseudo-chemical potential $\tilde{\mu}$ isotherm are displayed as a line for the $(T, V, N)$ data, and as  circles for the $(T, V, \mu)$ data. $L=40$ and $t=0.5$.}
\end{figure}

If the modified theory of Hill is fulfilled (even for finite systems), we expect the $(\rho, \tilde{\mu})$ values at the loops (Fig. \ref{figure3}) to be  given by the extrema of the distribution functions $P(N)$ of the grandcanonical
ensemble (Fig. \ref{figure2}).
In Fig \ref{figure4}, we compare function $\tilde{\mu}(\rho)$  obtained from the canonical ensemble simulations (Fig \ref{figure3}) with $\rho(\tilde{\mu})$ obtained from the extrema of the distribution
functions in the grand-canonical ensemble (Fig \ref{figure3}).  It can be seen that results coincide
entirely in the whole coexistence region. The coincidence of
both curves constitutes a good check on our modified version of Hill's theory.

An important and definite check on our proposal is possible. In the case of the 2d lattice
gas, an analytic expression for the interface tension was presented in Onsager's paper \cite{Onsager1944},
which, for the lattice interacting gas, is given by 
\begin{equation}
 \frac{\gamma}{\epsilon} = \frac{1}{2} - t \ln \left[ \frac{1+ e^{-\frac{1}{2t}}}{1-e^{-\frac{1}{2t}}}\right],
\end{equation}
with $t\equiv k_BT/\epsilon$.

Usually \cite{Troster, Binder2, Schrader} surface tension is calculated from the grandcanonical probability distribution function $P(N)$. In our proposal, numerical results for the function $\tilde{\mu}$ allow direct calculation of surface tension $\gamma$ from the
integration of Eq \ref{Equation10}. The interface has zero area at the gas density $\rho_{gas}$, increases as
density is increased, accompanying the formation of a spherical liquid bubble which
increases, deforms and becomes flat. As density is further increased, $\tilde{\mu}$ remains
constant, since $\left( \frac{\partial A_{int}}{\partial N}\right)_{T, V}=0$  as long as the interface remains flat, with $A_{int} = 2L^{d-1}$, in spite of the growth of the volume of the liquid phase. The integral over particle
number of Eq \ref{Equation4} may be written as 
\begin{equation}
 I \equiv \int_{\rho_{gas}}^{0.5}(\tilde{\mu}-\mu_{cx})d\left(\frac{N}{V}\right)= \gamma \int_{0}^{A_{plateau-int}} \frac{dA_{int}}{V}.
\end{equation}
From numerical integration, interface tension can be obtained
from $\gamma = \frac{I \times L}{2}$, for the case of the 2d lattice gas. 

Figure \ref{figure5} displays true and pseudo chemical potentials.The effect of interface area on the isotherms is substracted and true $\mu$ {\textit{vs}} density isotherms are obtained. The figure also illustrates our procedure for the obtention of surface tension. Fig \ref{figure6} shows our numerical results for $\gamma(t)$ as compared to the analytical prediction by Onsager \cite{Onsager1944}. 

\begin{figure}
\includegraphics[width=8.6cm, height=6cm]{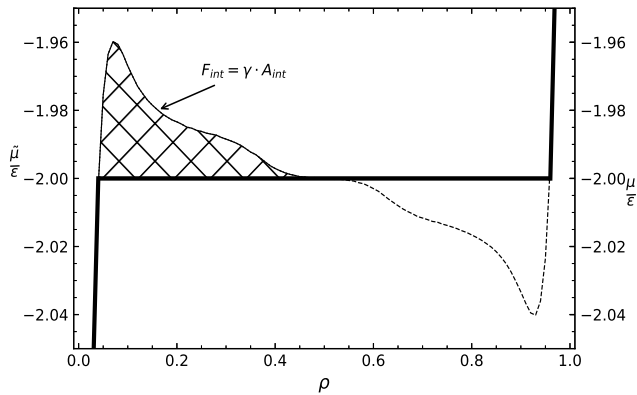}
 \caption{\label{figure5}True vs pseudo chemical potential isotherms. Extension of the plateau in $\tilde{\mu(\rho)}$ identifies limits of coexistence and gas and liquid densities $\rho$. The thick line represents the true chemical potential isotherm $\tilde{\mu(\rho)}$. Integration of $\tilde{\mu}$ positive loop allows for calculation of surface tension $\gamma (t)$. See text. Data are for $L=40$ and temperature $t=0.5$. }
\label{figure5}
\end{figure}

\begin{figure}
\centering
\includegraphics[width=8.8cm, height=6cm]{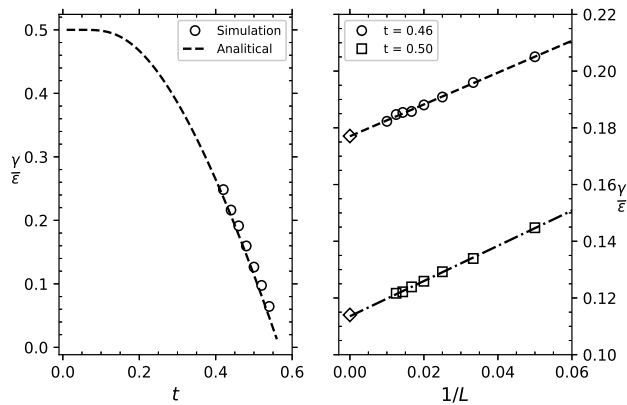}
 \caption{\label{figure6}Lattice gas surface tension $\gamma(t)$. In the left figure our $L=40$ data for $\gamma(t)$ are displayed as circles, and can be compared to Onsager's exact result (dashed line). The right figure displays behaviour with size $L$ for temperatures $t=0.45$ and $t=0.5$. The exact results, corresponding to $L \rightarrow \infty$ are indicated with the diamond symbol. For $t=0.46$, fitting yields $\gamma(t, L \rightarrow \infty) = 0.1135$, compared to exact $\gamma(t=0.46) = 0.1140$.  For $t=0.5$, fitting yields $\gamma(t, L \rightarrow \infty) = 0.1770$, compared to exact $\gamma(t=0.5) = 0.1771$.}
\end{figure}

In summary, 
we have found an interpretation for loops of thermodynamic fields encountered in
numerical simulations in the corresponding density ensembles. We have given proof
that the numerical results are not in fact results for the fields, but results for the sum of
the thermodynamic field and the gradient of the interface area that arises in the
coexistence region. The true thermodynamic fields maintain their proper convexity, 
and the numerical results in the phase coexistence region in the two ensembles are identical. As a side product of our analysis, we propose a much simpler
method for the calculation of surface tension, turning unnecessary the non-trivial
calculation of the full grand-canonical distribution probability function. 

\bibliography{apssamp}

\end{document}